\documentclass[prx,a4paper,aps,twocolumn,superscriptaddress,10pt,bibnotes]{revtex4-2}
\usepackage{graphicx,color}% Include figure files
\usepackage{dcolumn}% Align table columns on decimal point
\usepackage{bm}% bold math
\usepackage{amsmath,amssymb,mathrsfs}
\usepackage{url}
\usepackage{hyperref}
\usepackage{tikzit}
\tikzstyle{node}=[fill={rgb,255: red,64; green,64; blue,64}, draw=black, shape=circle]

\tikzstyle{dashed line}=[-, dashed]
\tikzstyle{red line}=[-, draw=red]
\tikzstyle{blue line}=[-, draw=blue, dashed]
\tikzstyle{arrow}=[->, >=latex]
\tikzstyle{double arrow}=[->, double, >=latex]
\tikzstyle{thick arrow}=[->, >=latex, thick]
\tikzstyle{red thick arrow}=[->, draw=red, thick]
\tikzstyle{blue thick arrow}=[->, draw=blue, thick]

\setcitestyle{sort&compress,numbers}

\def\>{\rangle}
\def\<{\langle}

\begin{document}
\title{Surpassing the Global Heisenberg Limit Using a High-efficiency Quantum Switch}
    
\author{Yu Guo}
\email{These two authors contributed equally to this work.}
\affiliation{Laboratory of Quantum Information, University of Science and Technology of China, Hefei 230026, China}
\affiliation{CAS Center For Excellence in Quantum Information and Quantum Physics, 	University of Science and Technology of China, Hefei 230026, China}
\affiliation{Anhui Province Key Laboratory of Quantum Network, University of Science and Technology of China, Hefei 230026, China}

\author{Yuehan Chen}
\email{These two authors contributed equally to this work.}
\affiliation{Laboratory of Quantum Information, University of Science and Technology of China, Hefei 230026, China}
\affiliation{CAS Center For Excellence in Quantum Information and Quantum Physics, 	University of Science and Technology of China, Hefei 230026, China}
\affiliation{Anhui Province Key Laboratory of Quantum Network, University of Science and Technology of China, Hefei 230026, China}

\author{Geng Chen}
\affiliation{Laboratory of Quantum Information, University of Science and Technology of China, Hefei 230026, China}
\affiliation{CAS Center For Excellence in Quantum Information and Quantum Physics, 	University of Science and Technology of China, Hefei 230026, China}
\affiliation{Anhui Province Key Laboratory of Quantum Network, University of Science and Technology of China, Hefei 230026, China}
\affiliation{Hefei National Laboratory, University of Science and Technology of China, Hefei 230088, China}

\author{Xiao-Min Hu}
\affiliation{Laboratory of Quantum Information, University of Science and Technology of China, Hefei 230026, China}
\affiliation{CAS Center For Excellence in Quantum Information and Quantum Physics, 	University of Science and Technology of China, Hefei 230026, China}
\affiliation{Anhui Province Key Laboratory of Quantum Network, University of Science and Technology of China, Hefei 230026, China}
\affiliation{Hefei National Laboratory, University of Science and Technology of China, Hefei 230088, China}

\author{Yun-Feng Huang}
\affiliation{Laboratory of Quantum Information, University of Science and Technology of China, Hefei 230026, China}
\affiliation{CAS Center For Excellence in Quantum Information and Quantum Physics, 	University of Science and Technology of China, Hefei 230026, China}
\affiliation{Anhui Province Key Laboratory of Quantum Network, University of Science and Technology of China, Hefei 230026, China}
\affiliation{Hefei National Laboratory, University of Science and Technology of China, Hefei 230088, China}
    
\author{Chuan-Feng Li}
\affiliation{Laboratory of Quantum Information, University of Science and Technology of China, Hefei 230026, China}
\affiliation{CAS Center For Excellence in Quantum Information and Quantum Physics, 	University of Science and Technology of China, Hefei 230026, China}
\affiliation{Anhui Province Key Laboratory of Quantum Network, University of Science and Technology of China, Hefei 230026, China}
\affiliation{Hefei National Laboratory, University of Science and Technology of China, Hefei 230088, China}

\author{Guang-Can Guo}
\affiliation{Laboratory of Quantum Information, University of Science and Technology of China, Hefei 230026, China}
\affiliation{CAS Center For Excellence in Quantum Information and Quantum Physics, 	University of Science and Technology of China, Hefei 230026, China}
\affiliation{Anhui Province Key Laboratory of Quantum Network, University of Science and Technology of China, Hefei 230026, China}
\affiliation{Hefei National Laboratory, University of Science and Technology of China, Hefei 230088, China}

\author{Bi-Heng Liu}
\email{bhliu@ustc.edu.cn}
\affiliation{Laboratory of Quantum Information, University of Science and Technology of China, Hefei 230026, China}
\affiliation{CAS Center For Excellence in Quantum Information and Quantum Physics, 	University of Science and Technology of China, Hefei 230026, China}
\affiliation{Anhui Province Key Laboratory of Quantum Network, University of Science and Technology of China, Hefei 230026, China}
\affiliation{Hefei National Laboratory, University of Science and Technology of China, Hefei 230088, China}

\begin{abstract}
Indefinite causal orders have been shown to enable a precision of $N^{-2}$ in quantum parameter estimation, where $N$ is the number of independent processes probed in an experiment. This surpasses the widely accepted ultimate quantum precision of the Heisenberg limit, $N^{-1}$. While a recent laboratory demonstration highlighted this phenomenon, its validity relies on postselection for it only accounted for a subset of the resources used. Achieving a true violation of the Heisenberg limit—considering photon loss, detection inefficiency, and other imperfections—remains an open challenge. Here, we present an ultrahigh-efficiency quantum switch to estimate the geometric phase associated with a pair of conjugate position and momentum displacements embedded in a superposition of causal orders. Our experimental data demonstrate precision surpassing the global Heisenberg limit without artificially correcting for losses or imperfections. This work paves the way for quantum metrology advantages under more general and realistic constraints.                
\end{abstract}

\maketitle
    % \tableofcontents
    % \newpage

%\section{Introduction}
Quantum metrology \cite{giovannetti2004quantum,giovannetti2006quantum,giovannetti2011advances}, a cornerstone of quantum information science, harnesses quantum resources to surpass classical precision limits in parameter estimation. In photonic systems, recent works have demonstrated precision scaling beyond the standard quantum limit (SQL, $N^{-1/2}$) through multiphoton entanglement\cite{nagata2007beating,dowling2008quantum,xiang2011entanglement} and multi-pass coherent interrogation \cite{higgins2007entanglement,juffmann2016multi,braun2018quantum}, approaching the theoretical Heisenberg limit (HL, $N^{-1}$). To address practical challenges of environmental decoherence, additional control\cite{tan2013enhancement,lang2015dynamical,yuan2015optimal,hou2019control} and learning based\cite{xu2019generalizable,zuo2022deep,xiao2022parameter} mechanisms have been implemented to dynamically optimize system evolution trajectories. These configurations effectively suppress technical noise and recover Heisenberg-level precision, significantly advancing the development of practical quantum-enhanced measurement devices.

Nevertheless, some discrepancies persist in assessing genuine quantum advantages within current protocols. Current experimental validations predominantly employ postselection techniques that compute quantum Fisher information~\cite{helstrom1969quantum} exclusively from surviving photons, systematically neglecting the photon loss inherent to real-world measurement processes. Consequently, the claimed quantum superiority over classical precision benchmarks becomes operationally ambiguous, as classical strategies must account for all initially prepared resources rather than postmeasurement survivors. Such inconsistencies necessitate a rigorous reassessment of the quantum metrological superiority relative to classical strategies. A critical breakthrough was recently achieved by Slussarenko {\it et al.}~\cite{slussarenko2017unconditional}, who demonstrated unconditional violation of the SQL using entangled photon pairs in a Mach-Zehnder interferometer, with precision exceeding classical bounds under full photon-loss accounting.

On the other hand, the emergence of indefinite causal order (ICO) has expanded the quantum information toolkit over the last decades\cite{chiribella2013quantum,oreshkov2012quantum}. Originally conceived in quantum gravity studies \cite{rovelli2004quantum,hardy2007quantumGravity}, ICO-enabled processes, exemplified by the quantum switch~\cite{chiribella2013quantum}, allow superposition of temporal sequences for physical operations. This paradigm has been shown to offer advantages in a wide variety of quantum-information processing tasks~\cite{chiribella2012perfect,Procopio2015Experimental,Araujo2014ComputationalAdvantage,taddei2021computational,Ebler2018EnhancedCommunication,Salek2018QuantumCommunication,goswami2020IncreasingCommunication,guo2020experimental,Rubino2021Communication,DJproblem,raz1999exponential,Guerin2016communicationComplexity,wei2019experimentalCommunication,felce2020QuantumRefrigeration,chen2021indefinite,Guha2020Thermodynamic,zhu2023prl,cao2022quantumSimulation,nei2022NMRswitch,Liu23thermodynamics,xi2024experimental,tang2024demonstration,zhao2019QuantumMetrology,Frey2019depolarizingChannelIdentification,Ban2023FisherInformation,Chapeau2021NoisyQuantumMetrology,yin2023experimental,Min23};  see Ref.~\cite{rozema2024experimental} for a review. In quantum metrology, Zhao {\it et al.}~\cite{zhao2019QuantumMetrology} reported that ICO-enhanced protocols could transcend the widely believed HL. They demonstrated that coherent control over 2N processes in infinite-dimensional systems enables geometric phase estimation with super-Heisenberg scaling ($N^{-2}$). Crucially, any fixed-order strategy remains confined to Heisenberg scaling. Experimental validation followed in 2023~\cite{yin2023experimental}, where a photonic quantum switch achieved super-Heisenberg precision in estimating geometric phases from transverse position-momentum displacements, albeit under idealized conditions.

\begin{figure}[htbp]
    \centering
    \includegraphics[width=0.9\linewidth]{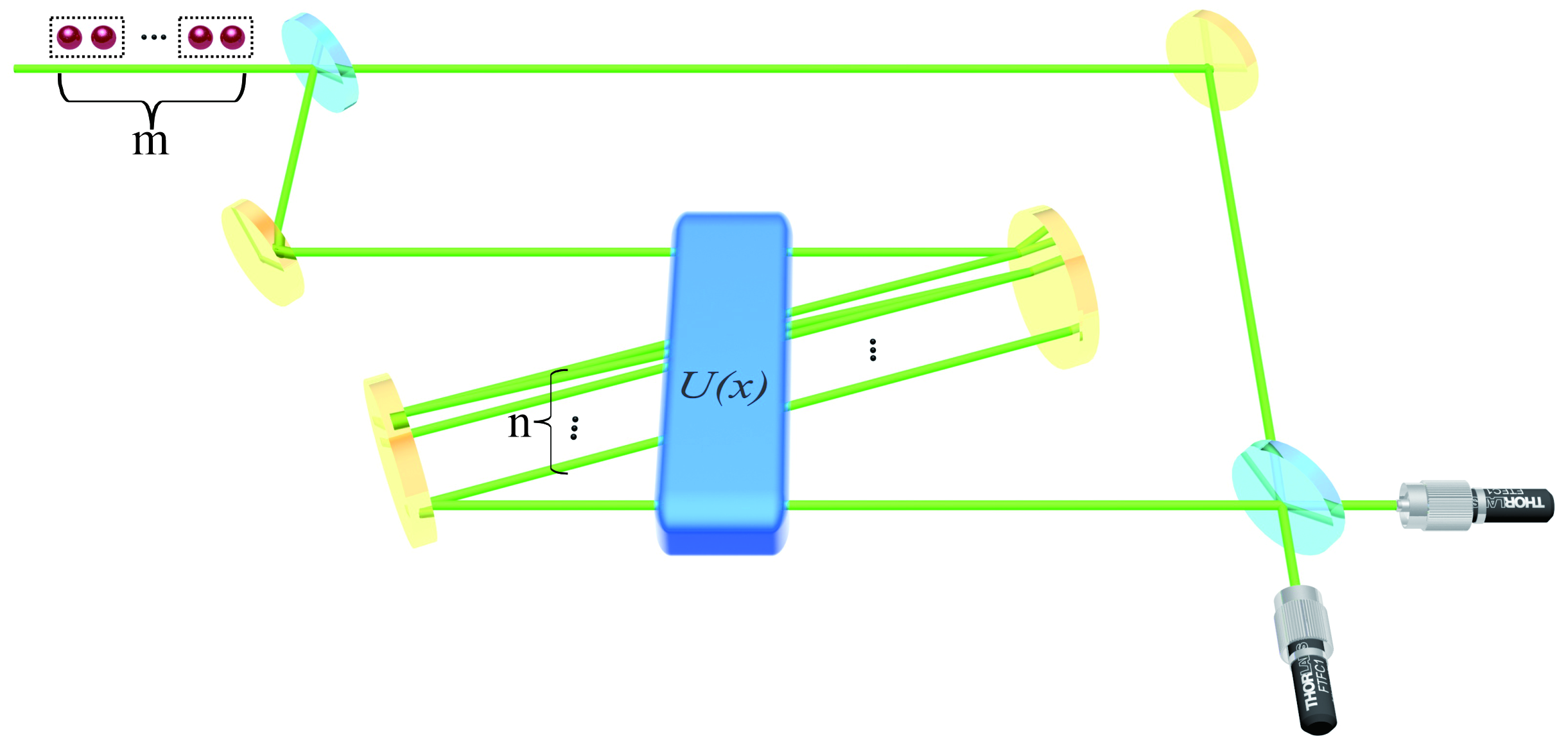}
    \caption{\emph{Quantum resource in Mach-Zenhder interferometer phase estimation.} In this scenario, $m$ photons are injected in the interferometer and each photon goes through $n$ passes of a phase shift $U(x)$. Parts or all of the incident photons can be entangled. The number of total quantum resource in this task is $mn$.}
    \label{fig:MN}
\end{figure} 

Here, we extend this line of research by investigating the conditions required to achieve a violation of the global HL in the ICO-enhanced metrology task, fully accounting for all experimental imperfections. Using a photonic platform, we demonstrate a geometric phase estimation precision that surpasses the upper bound achievable by ICO-free strategies with the same independent processes $mn$, where $m$ denotes the number of photons and $n$ the number of position-momentum displacement pairs. Specifically, we implement a high-efficiency quantum switch and design displacements within the photonic position-momentum phase space of its propagation mode. The geometric phase, corresponding to the area enclosed by two distinct orders of these displacements, is then estimated. In the experiment, $m = 60$ photons and up to $n = 30$ displacement pairs are employed for phase estimation and the photon source achieves a coincidence efficiency of approximately $71.5\%$, incorporating two superconducting nanowire detectors with efficiencies of $88.0\%$ and $89.0\%$, respectively. The experimental data demonstrate an unconditional violation of the global HL by at least $3$ standard deviations. This work establishes a new benchmark in quantum parameter estimation, paving the way for practical applications of ICO-based techniques under realistic experimental conditions. 

Note that the experimental challenge intensifies when pursuing postselection-free super-HL versus super-SNL. Taking the canonical model of phase estimation in a Mach-Zehnder interferometer as an example (see Fig.~\ref{fig:MN}). The total quantum resource is $mn$ with $m$ and $n$ being the number of photons used and the independent evolution each photon undergoes, respectively. The difficulty comes from the fact that one needs to repeat the experiment for enough rounds to calculate the precision but can hardly entangle all $m$ photons. So, only part of the photon can contribute inversely proportional to the precision which is harmful for the demonstration of the super-HL.

%%%%%%%%%%%%%%%%%%%%%%%%%%%%%%%%%%%%%%%%%%%%%%%%%%%% 
\begin{figure*}[htbp]
    \begin{center}
    \includegraphics[width=1.2\columnwidth]{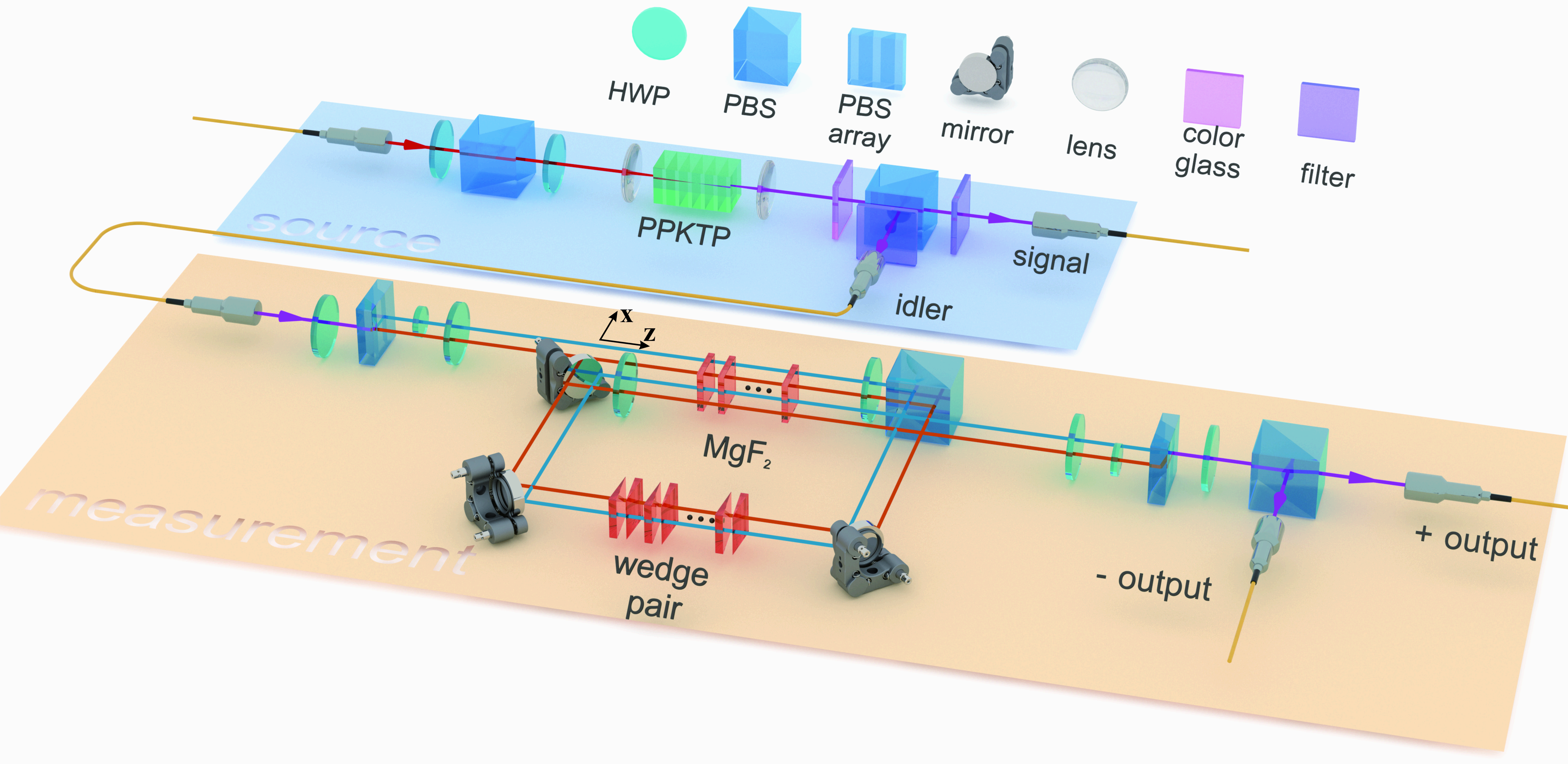}
    \end{center}
    \caption{\emph{Experimental setup.} A 20~mW cw violet laser at 775~nm pumps a type-II cut ppKTP crystal, effectively working  as a heralded single photon source when the signal photons trigger a single photon detector. To optimize the correlation-mode efficiency, the combination of the focusing lens are used to adjust the beam waists of both the pump mode and collection modes. In the quantum SWITCH, the idler photon's polarization serves as the control qubit and is initialized with a half wave plate (HWP). Transverse modes of the photon serve as the target system, which undergoes the position and momentum displacements $D_{x_j}$ and $D_{p_k}$. $D_{x_j}$ and $D_{p_k}$ are realized using magnesium fluoride (MgF$_2$) crystal and quartz wedge pair respectively and can be opened or closed using four HWPs. The control qubit is initialized in $|+\>=(|H\>+|V\>)/\sqrt{2}$ and this induces a superposition of paths through a Mach–Zehnder interferometer (between the two polarizing beam splitter arrays), with the two paths traversing $D_{n\bar{x}}$ and $D_{n\bar{p}}$ in opposite orders. Finally, the control qubit is measured using a HWP and a polarizing beam splitter (PBS) to estimate the geometrical phase $A$. %HWP, half wave plate; PBS, polarizing beam splitter; RM, reflection mirror; FC, fiber coupler; IF, interferometer filter.
    }
      \label{fig:experimentalsetup}
\end{figure*} 

%%%%%%%%%%%%%%%%%%%%%%%%%%%%%%%%%%%%%%%%%%%%%%%%%%%% 
\medskip
{\em ICO-enhanced quantum metrology with experimental imperfections.---} We begin by reviewing the framework of super-Heisenberg scaling geometric phase estimation through superposition of alternative causal orders, with particular emphasis on analyzing experimental imperfection effects within this protocol. Our framework considers $n$ pairs of black boxes, where each pair encodes either a position displacement $D_{x_j}=e^{-ix_jP}$ or a momentum displacement $D_{p_k}=e^{-ip_kX}$ (with $j,k=1,...,n$). Here, $X$ and $P$ represent conjugate variables defined through annihilation ($a$) and creation ($a^\dag$) operators as 
$X=(a+a^\dag)/\sqrt{2}$ and $P=i(a^\dag-a)/\sqrt{2}$. The primary objective of this protocol is to estimate the product $A=\bar{x}\bar{p}$, where $
\bar{x}$ and $\bar{p}$ denote the arithmetic means of the unknown displacement parameters $x_j$ and $p_k$, respectively.

It has been proven that, under energy-constrained scenarios, the root mean square error (RMSE) for estimating $A$ through fixed-order strategies (whether parallel or sequential) scales fundamentally limited by HL~\cite{zhao2019QuantumMetrology}. New possibilities arise when implementing quantum-controlled superposition of causal orders between the displacements. Specifically, by embedding the total displacements $D_{n\bar{x}}=\prod_{j=1}^nD_{x_j}$ and $D_{n\bar{p}}=\prod_{k=1}^nD_{p_k}$ within a quantum switch, with the control qubit initialized in the superposition state $|+\>=\frac{1}{\sqrt{2}}(|0\>+|1\>)$, we prepare the composite system state:
\begin{align}\label{SwitchedXP}
\frac{1}{\sqrt{2}}\left(|0\rangle\otimes \prod_{j=1}^nD_{x_j}\prod_{k=1}^nD_{p_k}|\psi\rangle + |1\rangle\otimes\prod_{k=1}^nD_{p_k}\prod_{j=1}^nD_{x_j}|\psi\rangle\right),
\end{align}
where $|\psi\>$ represents the target system's initial state. Fourier basis measurements $\{|+\>, |-\>\}$ on the control qubit yield outcome probabilities:
\begin{align}\label{ProSW}
P_{\pm}=\frac{1}{2}\left(1\pm \cos{n^2A}\right).
\end{align}
This measurement scheme directly generates Fisher information
\begin{align}
F_A=\sum_{o\in{\pm}}\frac{(\partial P_o/\partial A)^2}{P_o}=n^4,
\end{align}
establishing the ultimate precision limit $\delta A=\frac{1}{\sqrt{m}n^2}$, where $m$ denotes the number of experimental repetitions. Notably, the parameter $A$ corresponds to the geometric phase arising from $D_{x_j}$-$D_{p_k}$ sequences under alternative causal orders~\cite{yin2023experimental}.

This protocol demonstrates unambiguous super-Heisenberg scaling that remains effective in standard metrology configurations employing energy-conserving probes and independent interaction processes. Crucially, it differs fundamentally from super-HL schemes that rely on inefficient detectors~\cite{beltran2005breaking}, nonlinear coupling dynamics~\cite{boixo2007generalized,roy2008exponentially,napolitano2011interaction}, or probe-level non-Markovianity~\cite{yang2019memory}. Nevertheless, experimental realization of ICO-enhanced precision surpassing the HL via well-established photonic quantum switch implementations faces practical challenges, as decoherence mechanisms including photon loss, detector inefficiency, and reduced interferometric visibility introduce estimation uncertainties.

To quantify these effects, we model the main noise parameters: photon detection efficiency $\eta$ and interferometer visibility $\nu$. Our protocol can be mapped to the phase estimation scheme depicted in Fig.~\ref{fig:MN}. Incorporating visibility effects modifies the interference fringes to
\begin{align}\label{noiseProSW}
P_{\pm}^\prime=\frac{1}{2}\left(1\pm \nu\cos{n^2A}\right),
\end{align}
yielding the noise-affected Fisher information
\begin{align}
F_A^\prime=\frac{\nu^2n^4\sin^2{n^2A}}{1-\nu^2\cos^2{n^2A}}.
\end{align}
This leads to a precision of $\delta A^\prime=\frac{1}{\sqrt{mF_A^\prime}}$. Beating the HL requires $\delta A^\prime<\eta/mn$. At the point of minimum phase uncertainty, this reduce to 
\begin{align}\label{condition}
\frac{\eta^2\nu^2n^2}{m}>1.
\end{align}
Eq.~(\ref{condition}) establishes a criteria for unconditional violation of the HL via ICO-enhanced metrology, revealing that the repetition number $m$ plays a passive role in achieving this violation. This contrasts with the SQL violation criterion $\eta\nu^2n>1$ reported in Ref.~\cite{resch2007time}, where $m$ does not influence the boundary condition.

\medskip

{\em The experimental implementation.---} Our experimental setup, illustrated in Fig.~\ref{fig:experimentalsetup}, integrates a high-efficiency heralded single-photon source (blue region) with a hybrid-encoded quantum switch (orange region). Photon pairs are generated via type-II spontaneous parametric down-conversion (SPDC) in a 1-cm-long periodically poled KTP (ppKTP) crystal. The signal photon serves as a herald, while the idler photon is coupled into a single-mode fiber and directed into the quantum switch, where it undergoes $D_{x_j}$ and $D_{p_k}$ prior to detection for parameter $A$ estimation.

Achieving high photon detection efficiency is critical for our experiment. Losses arise not only from transmission and detection inefficiencies, but also from correlated-mode losses inherent to the SPDC process~\cite{dixon2014heralding,hu2022high}. To address this, we optimize the beam waists of both the pump and collection modes to enhance the correlated-mode efficiency. To minimize transmission loss, all optical components are coated with antireflection films at the corresponding wavelengths, and focus-adjustable optical couplers are used to match beam modes along the transmission path. When the waists of the pump and collection beams are set to $135 \mu m$ and $61 \mu m$, respectively, we observe a coincidence efficiency of $(71.5\pm 0.5)\%$, and at least $(50.6\pm 0.5)\%$ before and after the quantum switch. These values include the efficiency of the superconducting nanowire single-photon detectors (SNSPDs, $(88.5\pm 0.5)\%$). More details are provided in the Supplemental Material~\cite{SM}.

The control system of our quantum switch is the polarization qubit of the photon and its target system is defined as a continuous transverse mode perpendicular to the propagation direction $z$ ($x$-direction in Fig.~\ref{fig:experimentalsetup}), thus our setup presents a discrete- and continuous- hybrid encoded feature. The position displacement $D_{x_j}$ is generated with a customized 1.1-mm-thick birefringent MgF$_2$, which can translate the horizontally polarized photons by $x_j\approx9.125\mu m$ while transmitting the vertically polarized photons directly. As for the momentum displacement $D_{p_k}$, it is equivalent to a deflection of the photon with respect to the direction $z$ in the $x-z$ plane and can be achieved with a wedge pair~\cite{yin2023experimental}. In our experiment, one of the wedge is rotated by about $1^\circ$ and the photon is then deflected by $0.01^\circ$, thus generating a momentum displacement of $p_k\approx2\pi\theta_j/\lambda\approx709.4rad/m$ (in the paraxial approximation). These settings make sure that the total geometrical phase $A$ to vary within $2\pi$ with $n$ being up to 30. The tiny value of displacements, on the other hand, makes it difficult to calibrate the directions in our experiment. In Supplemental Material~\cite{SM}, we provide a method which enables to align the beams and the $z$ axis up to a precision of $\approx10^{-6}$ rad, which is two magnitudes smaller than $p_k$ we used. 

\begin{figure}[htbp]
    \centering
    \includegraphics[width=0.85\linewidth]{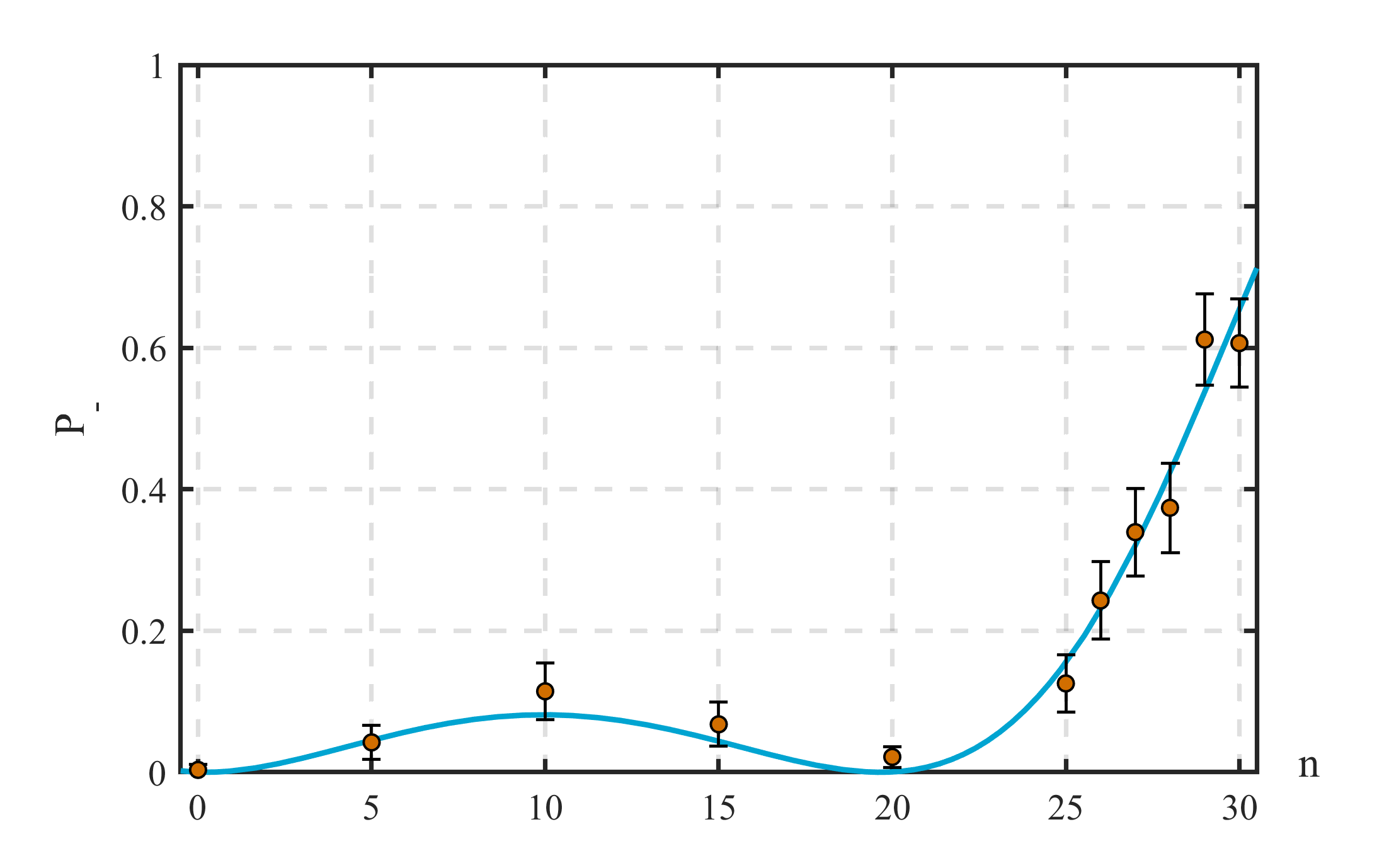}
    \caption{\emph{Probability  $P_-$ and the errorbar is  for $n\in[0, 30]$.} Our experimental data are marked as orange dots. The blue solid line is the fitting curve of these dots with the fitting function $\frac{1}{2}(1-\cos^2(An^2+cn+\phi_0))$. %the coefficients being $A = 0.00642$, $c = -0.130$, and $\phi_0 = 0.00833$.
    In each trial of experiment, $m=60$ photons are used to estimate the probability $P_-$ and the error bars (one standard deviation in this figure) are calculated through 30 trials of experiment.
    }
    \label{fig:pminus}
\end{figure}

In our experiment, the control qubit is initialized in superposition state $(|H\>+|V\>)/\sqrt{2}$, where $|H\>$ and $|V\>$ represent horizontal and vertical polarization states, respectively. Through a polarizing beam splitter (PBS) array (Fig.~\ref{fig:experimentalsetup}), these polarization states separate into distinct optical paths (blue/red) establishing coherent causal order superposition for displacement operations. The PBS array subsequently recombines both paths, followed by a Fourier measurement on the control qubit to estimate the values of $A$ and the corresponding precision.

We conducted the experiment using up to $n=30$ pairs of $D_{x_j}$ and $D_{p_k}$ and used $m=60$ counts to estimated the probability $P_-$ in each measurement trial. Our results are illustrated in Fig.~\ref{fig:pminus}, where the experimental values are marked with orange dots and are then fitted into the blue line. The fit value $A$ is 0.00642, which matches well with the design $A=0.00647$.
Note that there are two additional terms in the fitting function compared to the one in Eq.~(\ref{ProSW}). These terms result from the particular apparatus we used and will not contribute to the scaling of $A$'s precision~\cite{yin2023experimental} (see also Supplemental Material~\cite{SM}).

%The fitting function is $(1-\cos^2(An^2+cn+\phi_0))/2$, and extra phases $cn$ and $\phi_0$ appear compared to the function in Eq.~(\ref{ProSW}). The term $cn$ results from the use of the wedge pairs and the term $\phi_0$ is a fixed relative phase in the interferometer. \comnew{Also, the value of c can be calculate based on the wedge pair we used. The number is XX.} Note that these terms will not contribute to the scaling of $A$'s precision~\cite{yin2023experimental}. 

To demonstrate that our setup can violate the HL achievable by the global resource $mn/\eta$ without post-selection, we carefully examine the criterion in Eq.~(\ref{condition}) using our experimental parameters. In our implementation, we used $m=60$ photons, up to $n=30$ pairs of displacements, an interferometric visibility exceeding $\nu=0.989$, and a worst-case detection efficiency of $0.506$. In practice, an additional factor must be considered: the SPDC-based single-photon source occasionally generates multiple photon pairs with a small probability $\xi$. Ignoring this would lead to an underestimation of the total quantum resources consumed. In our experiment, $\xi$ was measured to be $0.04\%$ (see Supplemental Material~\cite{SM}). Therefore, when $m$ photons are detected, the actual number of photons used is $m(1+\xi)/\eta$. Under this condition, the revised criterion becomes $\frac{\eta^2\nu^2n^2}{m(1+\xi)^2}>1$. By substituting our experimental parameters, we obtain $\frac{\eta^2\nu^2n^2}{m(1+\xi)^2}\approx3.76$, which clearly exceeds the threshold.

\begin{figure}[thbp]
    \centering
    \includegraphics[width=0.95\linewidth]{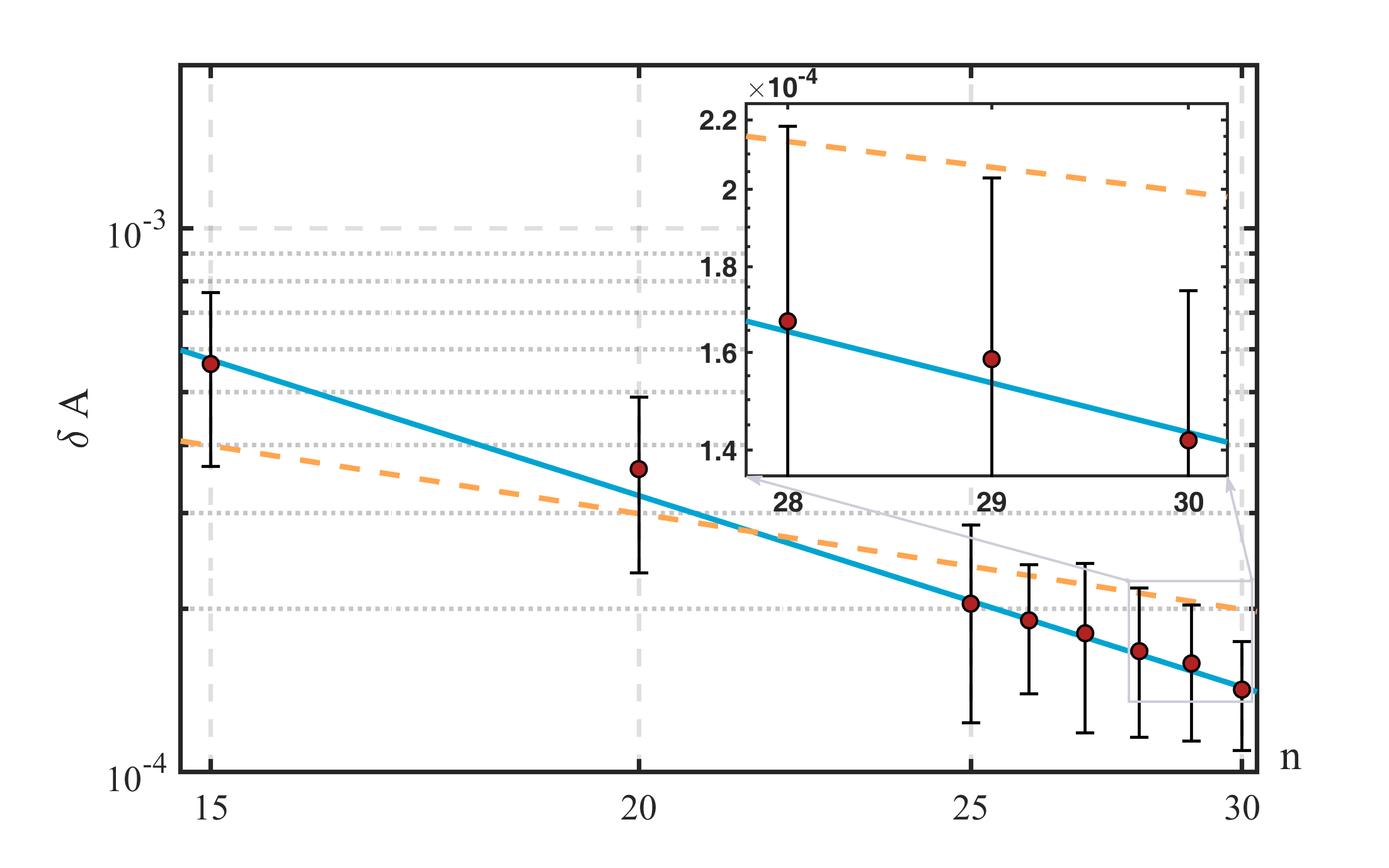}
    \caption{\emph{Super-HL precision of $A$ without artificially correcting experiment imperfections.} The orange dots describe the RMSE of $A$ from our experimental data; the blue solid line corresponds to the fitting result, which overlaps with the theoretical prediction $\frac{1}{\sqrt{\nu}N^2}$ (yellow solid line). By contrast, the ultimate precision limit of all fixed-order strategies using the actual resource consumed in our experiment is plotted in the yellow dashed line. The error bars ($3$ standard deviations in this figure) are calculated using bootstrapping method.}
    \label{fig:result1}
\end{figure}

We use the measured probability $P_-$ to estimate the geometric phase $A$ using the maximum likelihood estimation method. For each value of $n$, the experiment is repeated 30 times, and the resulting estimates of $A$ are used to calculate the RMSE $\delta A$, which quantifies the estimation precision. As illustrated in Fig.~\ref{fig:result1}, the experimental data (represented by orange dots) are fitted by a blue curve, while the theoretical HL corresponding to the actual resource consumed is shown as a orange dashed line. The fitting curve matches well with the theoretical prediction of $\frac{1}{\sqrt{\nu}N^2}$. The results show that, for $n>25$, the measured precisions lie below the HL, thereby confirming the unconditional violation of the global HL. To further quantify the statistical significance, we determine the standard deviation of $\delta A$ using bootstrap methods~\cite{davison1997bootstrap}. The analysis reveals that our precision surpasses the HL by at least three standard deviations when $n=29, 30$, thereby reinforcing the robustness of our violation.

\medskip  
%%%%%%%%%%%%%%%%%%%%%%%%%%%%%%%%%%%%%%%%%%%%%%%%%%%%  
{\em Discussion.---} In this paper, we have demonstrated an unconditional violation of global HL in an ICO-enhanced quantum metrology protocol. By integrating a high-efficiency heralded single-photon source with a hybrid-encoded quantum switch, we implemented geometric phase estimation via superposition of position-momentum displacement sequences. Crucially, our experiment achieves this super-Heisenberg scaling without post-selection, fully accounting for all photon losses and imperfections. The measured precision surpasses the HL attainable by any definite-order strategy using the same quantum resources, as confirmed by a $3\sigma$ statistical deviation. Our work marks a milestone for quantum estimation protocols, facilitating the application of ICO-enhanced methods in practical laboratory settings.

%\section{Data availability}
%The authors declare that the data supporting the findings of this study are available within the paper and in the supplementary information files.

%%%%%%%%%%%%%%%%%%%%%%%%%%%%%%%%%%%%%%%%%%%%%%%%%%%%

\section{Acknowledgments}
This work was supported by the NSFC (No.~12374338, No.~12174367, No.~12204458, and No.~12350006), the Innovation Program for Quantum Science and Technology (No.~2021ZD0301200), the Fundamental Research Funds for the Central Universities, Anhui Provincial Natural Science Foundation (No.~2408085JX002),  Anhui Province Science and Technology Innovation Project (No.~202423r06050004), China Postdoctoral Science Foundation (BX2021289 and 2021M700138),USTC Research Funds of the Double First-Class Initiative (No.~YD2030002026), the Ministry of Education Key Laboratory of Quantum Physics and Photonic Quantum Information (No. ZYGX2024K020).

%Y. G. and Y. C. contributed equally to this work.

\bibliography{references}

%\section{Author contributions}
%GC, BL, CL, and GG conceived the project and supervised the research. ZL developed the framework of witnesses of input-output indefiniteness and performed the numerical computations of the witnesses used in the experiments. BL and YG designed the experimental proposal. YG performed the experiment and analyzed the experimental data with the help of HT, XH, and YH.   ZL, YG and GC wrote the manuscript. YG and ZL contributed equally. 

%\section{Competing interests}
%The authors declare no competing interests.

%\newpage
%\appendix
%\onecolumngrid
%\begin{center}
%{\bf Experimental demonstration  of  input-output indefiniteness  in a single quantum device: Supplemental Information}
%\end{center}

\end{document}